\documentclass[seceq]{ptptex}
\usepackage{epsfig}

\def\lsim{\raise0.3ex\hbox{$<$\kern-0.75em\raise-1.1ex\hbox{$\sim$}}}
\def\gsim{\raise0.3ex\hbox{$>$\kern-0.75em\raise-1.1ex\hbox{$\sim$}}}

\title{
Dimensional reduction in QCD: Lessons from lower dimensions
}

\author{ 
Piotr \textsc{Bialas$^a$},
Andr\'e \textsc{Morel$^b$} and
Bengt \textsc{Petersson$\, ^c\,$}\thanks{Presented by B. Petersson}
}

\inst{ 
$^a$Inst. of Physics, Jagellonian University, Reymonta 4,  
33-059 Krakow, Poland \\
$^b$ Service de Physique Th\'eorique, CEA-Saclay, 91191 Gif-sur-Yvette
cedex, 
France \\
$^c$Fakult\"{a}t f\"{u}r Physik, Universit\"{a}t 
Bielefeld, D-33615 Bielefeld, Germany
}

\abst{
In this contribution we present the results of a series of investigations of 
dimensional reduction, applied to SU(3) gauge theory in 2 + 1 dimensions.
We review earlier results, present a new reduced model with $Z_3$ symmetry, 
and discuss the results of numerical simulations of this model.
}

\begin{document}

\maketitle

\section{Introduction}
Dimensional reduction is a powerful method to study the long distance 
behaviour of field theories at high temperature. It was first introduced in 
[1, 2].
It was further developed for gauge theories in [3, 4, 5].
The first quantitative studies using lattice simulations of the reduced model
were performed in [6, 7, 8, 9]. 
Quantitative results for QCD were further obtained in [10 - 15].
For a nice review see [16].

 In order to find the region of validity of dimensional reduction we have 
investigated in detail the reduction of SU(3) gauge theory in 2+1 dimensions
[17, 18, 19, 20, 21].
In section II we present the method, in Section III some of the results are
described. In Section IV an alternative form of dimensional reduction is 
 presented, which preserves the $Z_3$ symmetry of the action. 
Section V, finally, gives our conclusions.

\section{Perturbative dimensional reduction}
The properties of gauge theories at finite temperature is described by the 
partition function
\begin{equation}
Z = \int {\cal D} {\cal A}^a_\mu \, {\cal D} \psi^{(f)c} 
{\cal D}\bar\psi^{(f)c} e^{-S ({\cal A}_\mu , \psi , \bar\psi)}
\end{equation}
where ${\cal A}^a_\mu (x)$ are the gauge fields in the adjoint representation, 
and where we have included fermion fields of flavor $f$ in the 
fundamental representation of SU(3). The action $S$ at vanishing chemical 
potential is given by 
\begin{equation}
 S = \int^{1/T}_0 \, dx_0 \int d^{d_s} x \left[ \frac{1}{4} {\cal F}^a_{\mu\nu}
{\cal F}^a_{\mu\nu} + \sum_f \bar\psi^{(f)} (\not\!\!{\cal D} + m_f ) \psi^{(f)}
\right]
\end{equation}
The metric is Euclidean and $d_s$ is the number of space dimensions,
$x=(x_0 , \bar x)$. The 
physical case is $d_s = 3$. In this contribution we will mainly discuss a 
simpler theory, where the method of dimensional reduction can be tested in 
detail, namely pure SU(3) gauge theory in 2+1 dimensions 
$(d_s = 2)$ without fermion fields. 

The $x_0$-integration goes from zero to $1/T$, where $T$ is the 
temperature. For $T\rightarrow \infty$, and for distances $R\gg 1/T$ it 
seems plausible that one can integrate out the non static modes of the 
field perturbatively. It is important to note that the perturbation theory
of these modes does not have infrared divergences, in contrast to 
the perturbative expansion in the full theory. We thus obtain an effective model 
of the static modes in $d_s$ dimensions. Note that the fermion fields, which obey 
 Fermi statistics have antiperiodic boundary conditions in $x_0$, and  
 thus have no static modes. 
\begin{equation}
Z = \int {\cal D} {\cal A}^{st}_0 (\bar x) {\cal D} {\cal A}^{st}_i (\bar x) 
 e^{-S_{eff} ({\cal A}^{st}_0 (\bar x), {\cal A}^{st}_i (\bar x))}
\end{equation}
It can be shown, that for $T$ large and $R\gg 1/T$ $(|\bar p| \ll T)$, 
 $S_{eff}$ has a finite number of local terms at  given order in $g^2/T$,
$|\bar p | / T$ [5]. 
The coefficients are determined by the renormalized perturbative expansion, 
and there are thus no new free parameters. A pictorial description of 
 dimensional reduction is given in Fig. 1, where the word Higgs refers
 to the field ${\cal A}^{st}_0$.
Obviously the reduced theory demands much less computertime than the full 
theory. 

\vspace{1cm}

\begin{figure}[ht]
\centerline{
\epsfig{file=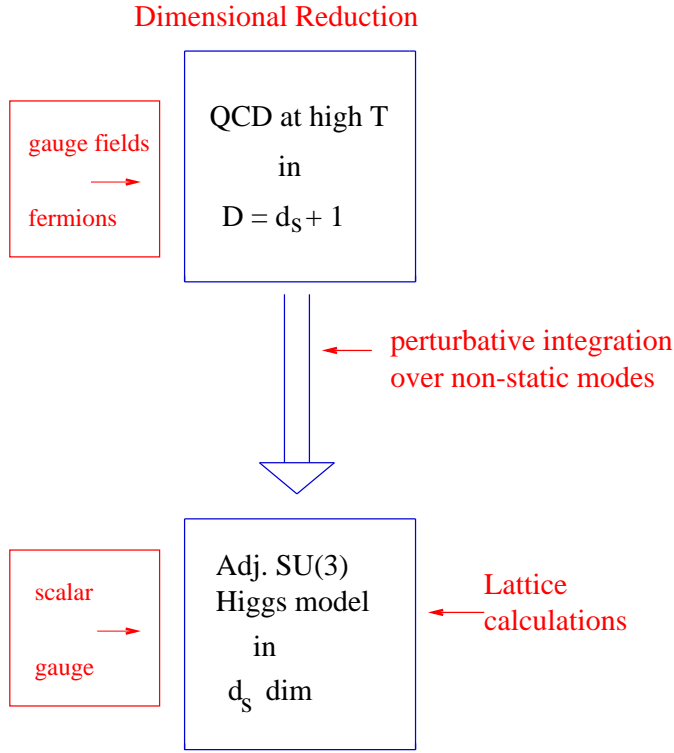,width=9cm}
}
\caption{
Pictorial description of dimensional reduction
}
\label{fig1}
\end{figure}

The static modes are defined by 
\begin{equation}
{\cal A}_\mu (x) = {\cal A}_\mu^{st} (\bar x) + {\cal A}^{ns}_\mu (x);\,\,
 \int^{1/T}_0 \, d x_0 {\cal A}_\mu^{ns} (x) = 0
\end{equation}
Note that this splitting is not gauge invariant in general. The physical 
quantities, which we calculate are, however, gauge invariant. 
Define
\begin{eqnarray}
{\cal A}^{st}_i (\bar x) & = & \sqrt T A_i (\bar x) \nonumber \\
{\cal A}^{st}_0 (\bar x) & = & \sqrt T \phi (\bar x) 
\end{eqnarray}
On the classical (tree) level we set 
\begin{equation}
 S_{eff} = \int d^{d_s} x \left[ \frac{1}{4} [F^a_{ij} (\bar x)]^2 +  
 \sum_{i=1}^{d_s} Tr [ D_i   \phi]^2 \right]
\end{equation}
This we also call ``naive reduction''. 
Because of the ultraviolet behaviour of the theory, we should, however, 
also take quantum effects into account. Restricting to the local terms, 
important for large distance physics, we find to one loop order the 
systematics described in Figure 2.

\medskip\noindent
Quantum theory \\
(one loop)
$$
\begin{array}{l l l}
& 3 + 1 \rightarrow 3  &~~~~~~   2 + 1 \rightarrow 2  \\
\epsfig{file=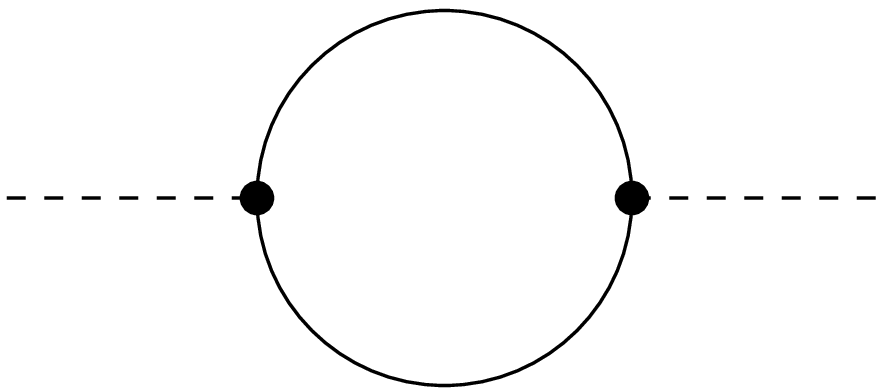, width=2.8cm}~~~~~~ & g^2 T^2~ \phi^2 &~~~~~~ 
g^2_3 T~\phi^2 \\
  & & \\
~\epsfig{file=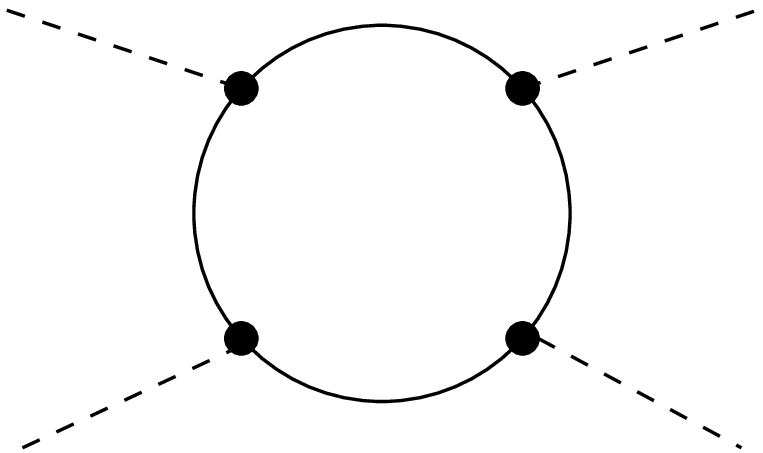, width=2.4cm}~~ & g^4 T~ \phi^4 &~~~~~~  
g^4_3~ \phi^4 \\
 & & \\
~\epsfig{file=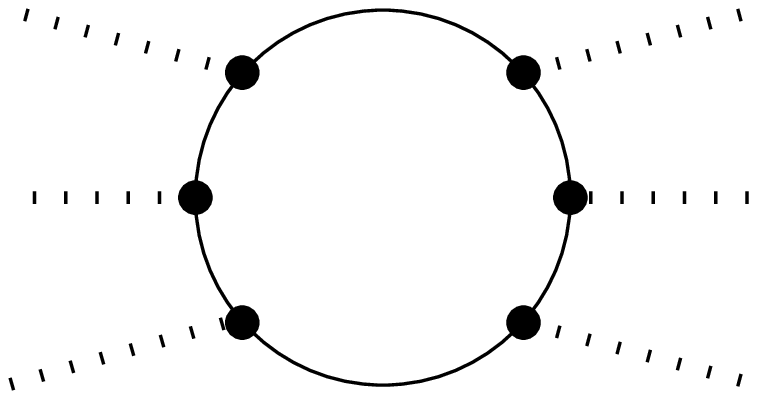, width=2.4cm} & g^6~ \phi^6  &~~~~~~
g^6_3/T~ \phi^6 \\
\end{array}
$$
\begin{figure}[h]
\caption{
Contributions to the $\phi^2, \phi^4$ and $\phi^6$ terms in the 
reduced action. In the diagrams, dotted lines are static modes, full
lines non-static modes. The numerical coefficients can be found in the
original papers, for 3+1 dimensions in [6, 7, 8, 10, 11] 
and for 2+1 dimensions in [17]. 
}
\label{fig2}
\end{figure}

\medskip\noindent
Note that in $d_s = 3$, $g^2$ is dimensionless. For $d_s=2$, $g^2_3$ has 
dimension energy, and $\phi$ is dimensionless. In the latter case, 
obviously higher loop orders are suppressed at high $T$ by powers of $g^2_3/T$.

The form of $S_{eff}$ calculated through perturbation theory of the 
non static modes in the lattice regularization can be found in 
[5, 6, 7, 8], and in dimensional regularization in [10] [11]. 
Here I present the result for $d_s = 2$, pure gauge theory, which we derived in 
[16]. 
We have chosen the static Landau gauge [STALG]
\begin{eqnarray}
\partial_0 {\cal A}_0 (x) &=& 0 \nonumber\\
\int^{1/T}_0 \partial_i {\cal A}_i (x) dx_0 &=& 0 
\end{eqnarray}
 We find to one-loop order in lattice regularization with spacing $a$
\begin{eqnarray}
S_2 \equiv S_{eff}& =& \int d^2 x \left[ \frac{1}{4} F_{ij}^a F_{ij}^a + 
\sum_i^{d_s} Tr [D_i   \phi]^2  \right. \nonumber \\
 & & \left.  -\frac{3g^2_3 T}{2\pi} [\frac{5}{2}\log 2 - 1 - \log aT]Tr \phi^2 + 
\frac{g^4_3}{32\pi} Tr \phi^4 \right]
\end{eqnarray}
The effective model is a Higgs gauge model in two dimensions, with the 
Higgs field $\phi$ in the adjoint representation. The action has a 
particular symmetry, $R$-symmetry, under $\phi \rightarrow -\phi$, which comes 
from the $T$-reflection symmetry in the 2+1 dimensional theory. 
Including fermions in the original model will only make a change of the value
of the coefficients. At finite density, however, $R$-symmetry and 
hermiticity is broken, and there are also terms with odd powers of $\phi$
and imaginary coefficients [15].

There is a logarithmic divergence in the coefficient of the quadratic 
term in $\phi$. This will be cancelled by the ultraviolet divergence of the 
two dimensional model. The perturbation theory of the model is infrared divergent.
 And although the pure gauge theory in two dimensions has been solved
     on the lattice [22], 
 and in the continuum in the large $N$ limit [23], 
there is no known solution for 2d gauge theory coupled to Higgs fields in the 
adjoint representation. Instead we solve the model numerically, using lattice
 Monte Carlo simulations. As a lattice version of this action we choose  
\begin{eqnarray}
S_{2L} & = & \beta_3 L_0 \sum_{\bar x}  \left( 1 - \frac{1}{3}
Re Tr U_p \right) \nonumber \\ 
& & + \sum_{\bar x , i} Tr \left(U (\bar x ; i) \phi (\bar x + a \hat i ) 
U (\bar x , i)^{-1} - \phi (\bar x )\right)^2  \\
 & & - \frac{9}{\pi L_0 \beta_3} \left[ \frac{5}{2} \log 2 - 1 + 
\log L_0 \right] Tr \phi^2 (\bar x ) + \frac{9}{8\pi \beta^2_3} Tr \phi^4 
\nonumber
\end{eqnarray}
The parameters $\beta_3$ and $L_0$ which appear in this action are related
to the coupling $g^2_3$ and the temperature $T$ of $QCD_{2+1}$ by
\begin{equation}
\beta_3 = \frac{6}{a g^2_3} ; \,\,\,\,\, L_0 = 1 / a T
\end{equation}
where $a$ is the lattice spacing. 

The results for physical quantities using this effective action are compared with
the full 2+1 dimensional SU(3) gauge theory, calculated with the lattice action
\begin{equation}
S_{3L} = \beta_3 \sum_x \sum_{\mu < \nu} \left( 1 - \frac{1}{3} Re Tr 
{\cal U}_p (x, \mu , \nu ) \right)   , 
\end{equation}
with the same spacing.

The matrices ${\cal U}_p$ and $U_p$ are the products of ${\cal U} (x , i)$
resp $U(\bar x , i)$ around 
a plaquette in 3 resp 2 dimensions, ${\cal U} ( x , i)$ and 
$U (\bar x , i) \in $ SU(3). 
The matrix $\phi (\bar x )$ is, however, in the algebra of SU(3); 
thus the $Z_3$-symmetry of $S_{3L}$ is broken in $S_{2L}$. 

 We     also define a dimensionless reduced temperature
\begin{equation}
\tau \equiv \frac{T}{g^2_3} = \frac{\beta_3}{6 L_0}  .
\end{equation}
The 2+1 dimensional model has a deconfining second order phase transition 
at [24]  
\begin{equation}
\tau = \tau_c = 0.614(3) .
\end{equation}

\section{Some results:}

Our main observable will be the correlation function of Polyakov loops 
$L (\bar x )$ and the corresponding screening mass i.e. inverse 
correlation length. 

In STALG we have 
\begin{eqnarray}
L ( \bar x ) &=& \frac{1}{3} Tr V(\bar x )  \\
V(\bar x ) &=& e^{i\phi (\bar x ) / \sqrt\tau}, \label{Vdef}
\end{eqnarray}
so that $L( \bar x )$ is a static operator and can be taken out of the 
integration over 
non-static variables. Therefore the comparison with the full model is 
straightforward. 


In Figure 3 we plot the screening mass as defined above. For details, see 
[17].
The screening masses in the naive tree level reduction do not agree with 
those of the full model. The agreement is, however, very good in the 
one-loop approximation form down to $T\approx 1.5 T_c$. At $T_c$ the screening
mass goes to zero in the full model. 

\begin{figure}[h]
\centerline{
\epsfig{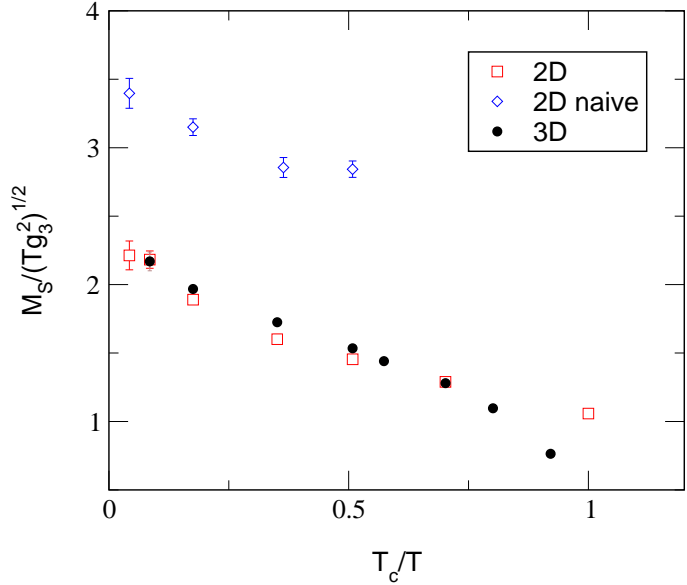}
}
\caption{
 Physical screening masses $M_S$ in units of $g_3 \sqrt T$ versus 
   $T_c / T$, in (2+1) $D$ (black points) and $2D$ (squares). Also shown are 
the masses obtained with the tree level reduced action (diamonds). 
}
\label{fig3}
\end{figure}
The reduced model does not have this second order phase transition, and thus 
there will be necessarily a discrepancy in the values of the screening masses
around $T = T_c$. 

We have also measured the spacelike string tension. Also here there is good
agreement between the full and the reduced model. In this case
at high temperature there is also good agreement with the analytic 
result of pure gauge theory in 2d.

In the 2d model one may in fact identify two screening masses, the 
lowest states $m_s$ of $Tr \phi^2$ $(R = 1)$ and $m_{ps}$ of 
$Tr \phi^3$ $(R = -1)$. 
The corresponding correlation functions are nearly straight exponentials, 
corresponding to well isolated poles. The ratio $m_{ps}/m_s$ between the two 
masses are between 1.5 and 2 depending on the temperature. Only a more 
detailed analysis could tell if there is agreement with the naive counting
rule $m_{ps} / m_s = 1.5$ [25]
For further details see [18].

The reduced model has two phases, a symmetric phase and a phase where 
the $R$-symmetry is spontaneously broken (Higgs phase). 
The transition in between is first order. The parameters in the one-loop 
approximation are in the unphysical broken phase, but very near 
the transition. In fact, we have performed the measurements at those values, 
but in the metastable symmetric phase region [17].

\section{$Z_3$-symmetric dimensional reduction}

To obtain a reduced $Z_3$ symmetric model 
we define instead of $U, \phi$ effective variables $U (\bar x , i)$ and 
$V (\bar x)$ which are SU(3) matrices, where 
$\frac{1}{3} Tr V (\bar x )$ is the Polyakov loop. 

Define
\begin{eqnarray}
Z_{eff} & = & \int d [U ] d [ V ] e^{-S_{eff}} \nonumber \\
S_U & = & \beta_3 L_0 \sum_{\bar x} \left( 1 - \frac{1}{3} Re Tr U_p \right) 
\nonumber \\
S_{U , V} & = & \frac{\beta_3}{L_0} \sum_{\bar x , i} \left( 1 - 
\frac{1}{3} Re Tr U (\bar x , i ) V ( \bar x + a \hat i) 
U (\bar x , i )^+ V (\bar x )^+ \right)  \\
S_V & = & \lambda_2 \sum_x | Tr V (x) / 3 |^2 \nonumber \\
S_{eff} & = & S_U + S_{U , V} + S_V \nonumber 
\end{eqnarray}
 The term $S_{U,V}$ is constructed such that developing $V (\bar x)$ in
 $\phi$ we get
in lowest order the kinetic term of $S_{2L}$ above. Similarly, developing 
$S_V$ we obtain terms proportional to $Tr \phi^2$ and higher orders. 

We investigated this model numerically, keeping $\lambda_2$ a 
free parameter [20, 21]. 
Similar models have been proposed in Refs [26, 27] 
 and for imaginary chemical potential in Refs [28, 29]. 

At sufficiently high temperature in the deconfined phase this model should 
be similar to the perturbative reduction. It has, however, an explicit 
$Z_3$-symmetry in the action, which may be broken at high temperature, but  
becomes restored at lower $T$. We have performed numerical simulations of the 
model, keeping $L_0 = 4$, and scanning the $(\beta_3 , \lambda_2)$ plane. 
With $L_0$ fixed, the temperature is proportional to $\beta_3$. To 
characterize the phases we use the distribution of the Polyakov loop. 

\begin{figure}[ht]
\centerline{
\epsfig{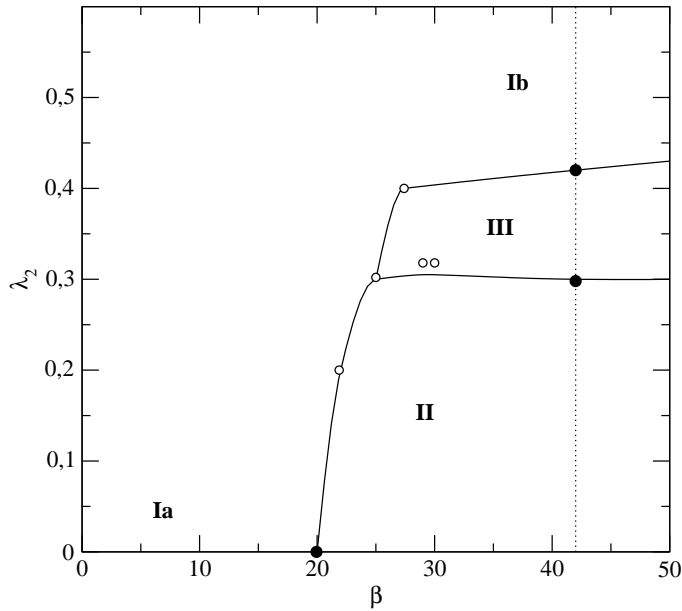}
}
\caption{
A tentative phase diagram. Open points denote the location of peaks in 
susceptibility observed on $16\times16$ lattices. Black points denote the 
position of phase transitions established on $32\times32$ lattices with big 
statistics. The lines are hand drawn sketches of where transitions take place.
}
\label{fig4}
\end{figure}
We find a quite non-trivial phase structure, as shown in Figure 4. 
For $\lambda_2 = 0$ there is in fact a transition to a confining phase at 
$\beta_3 = 20$, corresponding to $\tau_c = 0.83$, about 25\% higher than in the 
full model.  We checked that the screening mass actually vanishes
there, as it should. Then fixing $\beta_3$ at 42 $(\tau \approx 2 \tau_c)$, we made a more detailed 
investigation varying $\lambda_2$. The results for the 
susceptibility are plotted in Fig. 5.

\begin{figure}[h]
\centerline{
\epsfig{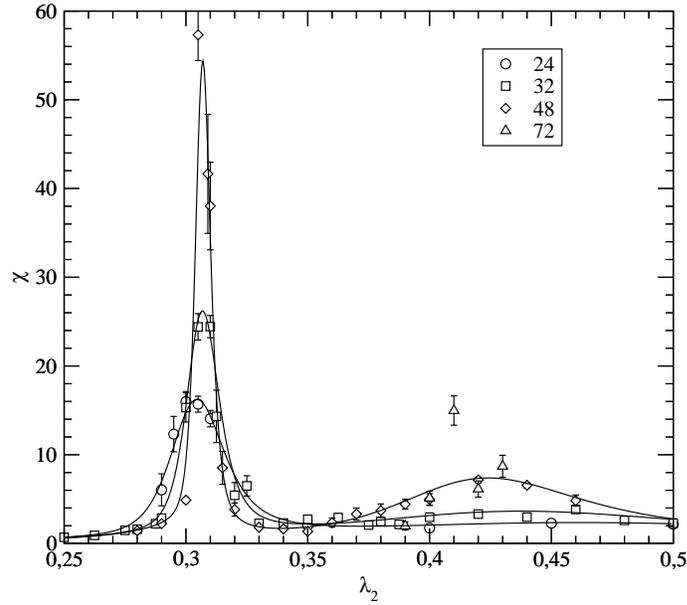}
}
\caption{
Polyakov loop susceptibility as a function of $\lambda_2$ for $\beta_3=42$ on  
 lattices of various sizes. The continuous lines were obtained through 
 Swendsen--Ferrenberg reweightings.
}
\label{fig5}
\end{figure}
We find in fact two transitions, corresponding to three different phases. In 
the phase II of Fig. 4
with small $\lambda_2$, large $\beta_3$, the values of $L$ are complex and
concentrated
near $(1, 0)$ up to $Z_3$ rotations, i.e. their phases are close to
$2i\pi n/3$. This is the normal
plasma phase. In the middle phase (phase III) $L$ has also 
non vanishing values near the border of phase space, but  
with phases close to $2i\pi (n+1/2)/3$. This is a new phase. Finally for large 
 $\lambda_2$, $L \approx 0$,  as it is in the deconfined phase, and 
we propose that this phase (Ib) is connected to the low temperature phase (Ia). 

It would be interesting to investigate, if 
the new phase plays a role in finite density QCD. 

We have investigated the behaviour of the screening mass. 
It is shown in Figure 6.  
Since here $\lambda_2$ is a free parameter, we cannot predict the screening 
mass from the reduced model. If we demand on the other hand that the 
 screening mass in the full and reduced models should be the same, we find
this to happen at $\lambda_2 \approx 0.18$, well inside the physical 
phase II, in contrast to the earlier method.

\begin{figure}[h]
\centerline{
\epsfig{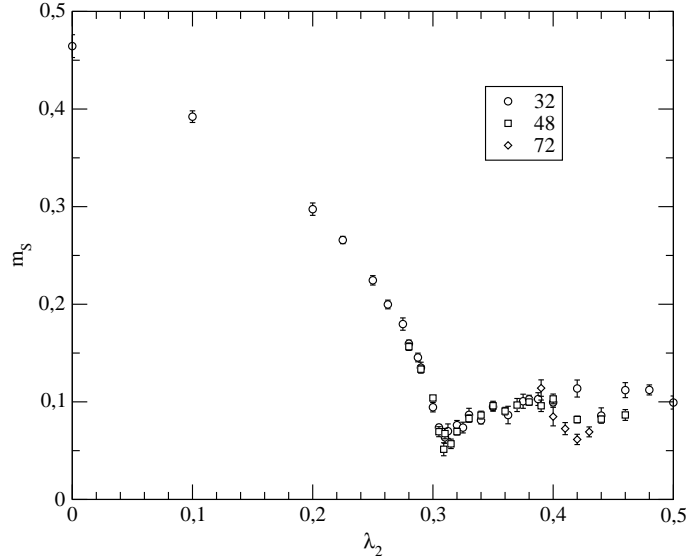}
}
\caption{
The screening masses $m_S$ for $\beta_3=42$ and various lattice sizes.
}
\label{fig6}
\end{figure}

We have also supplemented the numerical investigation with an analytic 
mean field calculation, which gives qualitatively the same results, in 
particular the phase structure. This calculation can easily be generalized
to the 3+1 dimensional case. Again one observes the same phase structure. 
For further details on the $Z_3$ symmetric reduction see [20, 21].

\section{Conclusions}
The usual method may be called perturbative reduction, although the reduced
theory is calculated non perturbatively. In this method there is a systematic 
expansion in $g^2/T$ and $\frac{1}{RT}$, where $R$ is the typical distance of 
interest. No free parameters are introduced. As we have shown, for 
finite temperature SU(3) gauge theory in 2+1 dimension this reduction works
very well for screening masses and spacelike string tension down to 
$T \approx 1.5 T_c$. 
A difficulty of principle is that the parameters determined from the 
reduction are not in the stable physical phase of the two dimensional model
but in the metastable physical phase slightly beyond the transition. 
Furthermore in this method, as $Z_3$ symmetry is explicitly broken, there is no 
confining transition. 

The $Z_3$ symmetric reduction, which we propose, works all the way down to 
the transition, and the parameters can be chosen in the physical phase. 

However, in this case we have not fixed the parameters of the reduced 
model from the full model. In order to keep the quantitative predictive power, 
this has to be done. 

Qualitatively the reduced model is in good agreement with the full model. 
We have also found a $Z_3$ symmetric new phase, which may be of importance 
at finite density.

\section*{Acknowledgments}

We are grateful to K. Petrov and T. Reisz, who were involved in the first part of this 
investigation. We thank ``Deutsche Forschungsgemeinschaft'' for support under 
the projekt FOR 339/2-1.
 BP thanks the Service de Physique Th\'eorique, CEA-Saclay, where part of this 
investigation was performed, for support and kind hospitality.

\end{document}